\documentclass[12pt]{iopart}
\usepackage{epsf}
\begin{document}

\title{Competing associations in six-species predator-prey models}
\author{Gy\"orgy Szab\'o}
\address {Research Institute for Technical Physics and
Materials Science, P.O. Box 49, H-1525 Budapest, Hungary}

\date{\today}

\begin{abstract}
We study a set of six-species ecological models where each species
has two predators and two prey. On a square lattice the time
evolution is governed by iterated invasions between the
neighboring predator-prey pairs chosen at random and by a site
exchange with a probability $X_s$ between the neutral pairs. These
models involve the possibility of spontaneous formation of
different defensive alliances whose members protect each other
from the external invaders. The Monte Carlo simulations show a
surprisingly rich variety of the stable spatial distributions of
species and subsequent phase transitions when tuning the control
parameter $X_s$. These very simple models are able to demonstrate
that the competition between these associations influences their
composition. Sometimes the dominant association is developed via a
domain growth. In other cases larger and larger invasion processes
precede the prevalence of one of the stable associations. Under
some conditions the survival of all the species can be maintained
by the cyclic dominance occurring between these associations.
\end{abstract}




\section{Introduction}
\label{sec:int}

Ecological systems exhibit a large number of associations of
certain species which coexist on the same territory. These
associations can be considered as evolutionary stable states (or
Nash equilibria) whose stability and competitiveness depend on the
environmental conditions
\cite{maynard:82,weibull:95,hofbauer:98,gintis:00,drossel:ap01,
miekisz:jpa04}. Generally, the systematic analysis of the
competition between these spatial states is made difficult by the
complexity of the phenomena and by the large number of parameters
in the suitable models.

In order to have a deeper insight into these phenomena we consider
now a family of toy  models exemplifying some basic features and
the richness of possible behaviors. In these predator-prey models
the species, located on the sites of a lattice, invades one of the
neighboring sites if it is occupied by its prey. The investigation of predator-prey systems on lattices has a long history reviewed recently in the papers \cite{drossel:ap01,pekalski_cse04,he_ijmp05}. In these systems the invasion
processes can yield extinction of species and/or maintain some
steady states in the spatial distribution if the corresponding
food web (a directed graph) contains loops. In such a situation a
self-organizing spatio-temporal pattern can be formed by the
species dominating cyclically each other. Besides it, in the
present model the neutral (neighboring) species are allowed to
exchange their sites. This latter process helps the formation of
many well-mixed distributions of neutral species which can also
protect each other against the rest of species. As a result these
models possess many final states toward which such a system can
evolve. In the spatial systems these states can coexist by forming
domains if the system is started from a random initial state for
sufficiently large sizes. The competition between these
multi-species associations along the boundaries can affect their
internal structure and the evolutionary process selects one of
these states. Thus, the investigation of these simplified models
can help us to clarify some mechanisms related to Darwinian
selection \cite{maynard:82} and the emergence of structural
complexity \cite{watt:je47} characterizing {\it
pre-biotic} evolution too \cite{rasmussen:science04}.

Several basic features are already well known for the simplest versions
of the above models. As mentioned above, if the spatial predator-prey
model on the square lattice contains three species with cyclic dominance
(this model is also called as spatial rock-scissors-paper game) then
the cyclic invasions maintain a self-organizing pattern with equal
species concentrations
\cite{tainaka:prl89,sato:mmit97,frachebourg:jpa98}. Both the composition
\cite{tainaka:pla93,frean:prs01} and the geometrical features
\cite{tainaka:epl91,szabo:pre02a} of this spatial pattern are affected
by the independent variation of invasion rates meanwhile this structure
provides a stability against some external invaders
\cite{boerlijst:pd91,szabo:pre01}. Similar self-organizing
pattern is found for those cyclic predator-prey (or voter)
models where the number $N_s$ of species is limited, namely $3 < N_s
\le 14$ \cite{frachebourg:jpa98}. For even $N_s$, however, these
systems are very sensitive to the independent variation of invasion
rates \cite{sato:amc02}. More complex behavior was
observed for those systems where the food web consists of several
cycles that can be distinguished from a topological point of view
\cite{szabo:pre01}. In these multi-cycle ecological
models the final (stationary) state is prevailed by a defensive alliance
being the most stable cyclic subsystem. These defensive alliances are
composed of those species that form a cycle in the food web thereby
their spatial distribution is equivalent to those of one-cycle systems
mentioned above. The privileged role of these associations is related to
their most effective protection, provided by the topological features of
the food web, against the external species. In their enhanced
protection the role of spatial structure is crucial because under
mean-field conditions these associations are not able to eliminate
the external invaders.

Very recently the investigation of a four-species cyclic predator-prey
model has indicated the spontaneous formation of another type of
defensive alliances \cite{szabo:pre04a}. In this latter model the
lattice sites can be empty and the individuals are allowed to jump to
one of the neighboring empty sites. As a result the odd (even)
label species can form a well-mixed state ensuring the possibility
of prompt counter-attack against the even (odd) label species and
by this means these associations can defend their territories. In other
words, the neutral species can also form defensive alliances in the
presence of some local mixing.

In the present paper we consider a set of models allowing the
appearance of both types of above mentioned defensive alliances.
In these six-species models the food web contains many cycles and
the local mixing is provided by a site exchange between the
neutral pairs. For all the models the probability $X_s$ of site
exchange will be the only tunable parameter. Using Monte Carlo
(MC) simulations it will be shown that, despite their simplicity,
these models are able to account different stationary states and
subsequent phase transitions when varying the value of $X_s$. The
stationary states are composed of different species whose number
varies from 2 to 6. Evidently, the time scales of the formation of
these states are dependent on the microscopic mechanism (invasion
or mixing). Sometimes equivalent states are formed via a
nucleation mechanism and they can destroy each other (cyclically)
when the boundaries of these growing domains meet.

Apparently the combination of our previous models
\cite{szabo:pre01,szabo:pre04a} means a simple extension of the
six-species models. This extension, however, exemplifies that a
small variation of a sufficiently complex ecological model can
yield surprisingly large increase in the types of behaviors and
phase transitions whose clarification raises many other questions.
At the same time, many aspects of these features seem to be valid
for the sufficiently complex ecological systems (or evolutionary
games) as discussed in the Conclusions.

\section{Spatial predator-prey models}
\label{sec:sppm}

\begin{figure}[h]
\begin{center}
             \epsfxsize=7cm
             \epsfbox{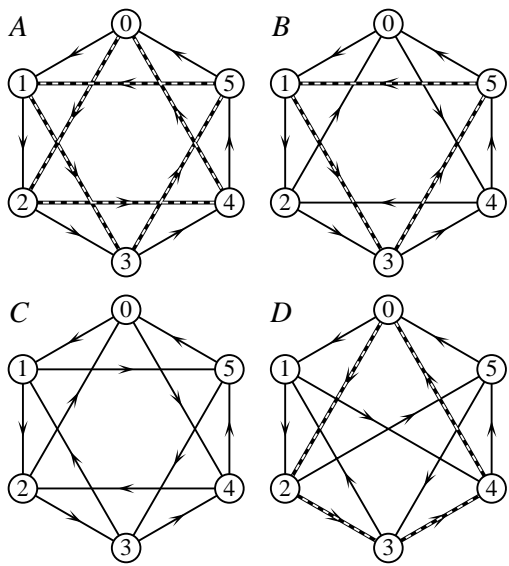}
\end{center}
\caption{Food webs defining the four classes of
predator-prey relations where each species has two predators and
two prey. The decorated lines connect the members of the
defensive alliances providing mutual protection against the
external invaders via cyclic invasions.}
\label{fig:foodweb}
\end{figure}

In all the present models the site $i$ of a square lattice is
occupied by an individual belonging to one of the six species and the
corresponding distribution is given by a set of site variables
($s_i=0, \ldots , 5$). The predator-prey relation between the
species is defined by a food web. On the corresponding directed
graph the nodes (labelled from 0 to 5) refer to the species and
the directed edges point to the prey from its predator. Our
analysis is focused on those systems where each species has two
predators, two prey, and a neutral one (not linked). In this case
each possible food web (satisfying the above conditions) is
isomorphic to one of the four graphs shown in
Fig.~\ref{fig:foodweb} and the corresponding model is named after
their labels A, B, C, and D \cite{szabo:pre01}.

The evolution of the species distribution is governed by iterating
nearest neighbor invasions and site exchanges. More precisely,
starting from a random initial state we repeat the following steps:
1) two nearest-neighbor sites are chosen at random; 2) if these sites are
occupied by a predator-prey pair then the prey is killed and an offspring
of the predator occupies the site of prey; 3) for neutral pairs the
species exchange their sites with a probability $X_s$. Evidently, nothing
happens if the individuals belong to the same species. For example, in
the model A the randomly chosen pair $(1,2)$ transforms into $(1,1)$
(invasion) and the pair $(1,4)$ transform to $(4,1)$ (site exchange)
with a probability $X_s$, otherwise the neutral pair remains unchanged.
Notice that the invasion rates between the predator-prey pairs are
unique and the models have only one parameter ($X_s$) characterizing
the strength of mixing between the neutral species.

These models are investigated by a series of MC simulations performed
on a square lattice with sites $N=L \times L$ under periodic boundary
conditions at different values of $X_s$. The simulations are started
from a random initial state. Within a time unit (MCS) each pair has a
chance once on the average to perform an invasion (between predator
and prey) or a site exchange with a probability $X_s$ (between neutral
species). During the simulations we have recorded the current values
of species concentrations [$\rho_{\alpha}(t)$; $\alpha = 0, \ldots, 5$]
as well as the probability of finding predator-prey [$P_{pp}(t)$] and
neutral pairs $P_{n}(t)$ on two neighboring sites. The stationary states
are characterized by the averages values of these quantities (denoted
as $\rho_{\alpha}$, $P_{pp}$, and $P_n$) which are determined by
averaging over a long sampling time varied from $10^4$ to $10^6$ MCS
after a suitable thermalization time. Evidently, $\sum_{\alpha}
\rho_{\alpha} =1$ and the quantity $1-P_{pp}-P_{n}$ gives the
probability of finding the same species on two neighboring sites.
At the same time we have determined the fluctuation of these quantities
defined as $\chi_{\alpha}= N [ \langle \rho_{\alpha}^2 \rangle -
\langle \rho_{\alpha} \rangle ^2]$ where $\langle \ldots \rangle$
refers to the averaging over the sampling time. To avoid undesired
size effects the linear size is varied from $L=500$ to 2500. The larger
sizes are used in the close vicinity of transition points where the
fluctuations diverge.

Now we discuss the general features of the possible stationary states.
For all these models there exist
six homogeneous states denoted as $\Phi_{\alpha}^{(h)}$ where
$s_i=\alpha$ ($\alpha = 0, \ldots , 5$) for each site $i$.
These states remain unchanged and the evolution of the species
distribution is stopped whenever one of these (absorbing) states is
reached. At the same time, the homogeneous state is unstable against
the invasion of the suitable predators.

The mixed states of neutral species are denoted as $\Phi_{03}^{(n)}$
if $s_i=0$ or 3. Evidently there exist two additional neutral pairs for
all the four models. In such a state there are no invasions at all.
Consequently, the ratio of populations is fixed during the evolution
meanwhile the system tends towards an uncorrelated spatial distribution
for $X_s>0$.

For all the four models there exist several three-species associations
whose members invade cyclically each other. For example, in model $A$
one of the the corresponding states is denoted as $\Phi_{024}^{(c)}$
where the subscripts refer to the three species being present with
the same average concentrations, $\rho_0=\rho_2=\rho_4=1/3$. The
corresponding spatial distribution is maintained by the cyclic
invasions yielding a short average life time for the individuals
(the survival probability of individuals decreases exponentially with
a relaxation time $\tau \simeq 1.8$ MCS) and the correlation length
($\xi \simeq 2.5$) is characterizing the typical domain size
\cite{tainaka:prl89,frachebourg:jpa98,ravasz:pre04}.
In general, the stability of the possible three-species associations
against the external invaders is determined by the food web
topology \cite{szabo:pre01}. For the model $A$ the food web contains
only two equivalent three-species cycles which are called defensive
alliances because their members protect cyclically each other against
the external invaders. For example, within the state $\Phi_{024}^{(c)}$
the species $2$ can only be invaded by the ``external'' species $1$.
During the cyclic invasion processes both the internal species $2$ and
the substituted invader $1$ are killed by their common predator ($0$),
in the meantime the third member ($4$) of the association feeds the
species $2$ and blocks the spreading of invaders. Thus, sooner or later
the offspring of invaders will be eliminated from the system. One
can easily check that other types ($3$ or $5$) of invaders become extinct
in the same way because of the cyclic symmetry in this food web.
For the food web $B$ the system has only one cyclic defensive alliance
($\Phi_{135}^{(c)}$) that will dominate the final stationary state
although this food web has four additional three-species cycles with
lower stability. Model $C$ has eight equivalent three-species cycles,
and neither of them can be considered as a defensive alliance in the
sense mentioned above.

Model $D$ represents another situation where one can find four
three-species and three four-species cycles. In this case the
four-species cycle $\Phi_{0234}^{(c)}$ proved to be the most stable
spatial association that satisfies the conditions to be a cyclic
defensive alliance \cite{szabo:pre01}.

For $X_s=0$, the stability of the different states is already
determined in a previous study when considering the effect of
mutations on the population in similar six-species predator-prey
models \cite{szabo:pre01}. The present analysis will be focused
on the effect of site exchange between the neutral pairs ($X_s>0$)
for the absence of mutation. For small sizes all the above
mentioned states can be observed as a final state. For
sufficiently large system sizes, however, the system tends towards
the real stationary state independent of the initial state.

\section{Four-species subsystems}
\label{sec:4s}

\begin{figure}[h]
\begin{center}
             \epsfxsize=7cm
             \epsfbox{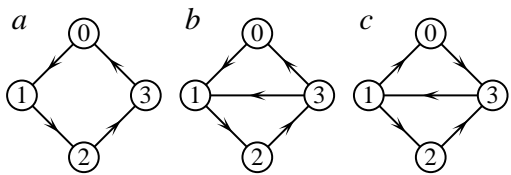}
\end{center}
\caption{Food webs for the four-species subsystems.}
\label{fig:foodwsub}
\end{figure}

In this section we study the behavior of several four-species
subsystems whose understanding will help the interpretation of
results obtained for the more complicated systems. More precisely,
our analysis is focused on three systems (labelled as $a$, $b$,
and $c$) whose food web are given in Fig.~\ref{fig:foodwsub}.

The food web of subsystem $a$ is equivalent to a four-species
cyclic predator-prey model where each species has only one
predator and one prey. Such a situation can occur in the models
$A$, $B$, and $D$ if two species are missing, and its analysis
becomes particularly important for model $D$ where
$\Phi_{0234}^{(c)}$ is the stable state for $X_s=0$. In addition,
similar cyclic dominance has been reported very recently by
Traulsen {\it et al.} \cite{traulsen:pre03,traulsen:pre04}
considering a four-strategy prisoner's dilemma game.

Some features of this system is already investigated previously by
several authors for $X_s=0$
\cite{frachebourg:jpa98,sato:amc02,szabo:pre04a}. In this case the
cyclic invasions maintain a self-organizing pattern with
$\rho_0=\rho_1=\rho_2=\rho_3=1/4$ and with a probability
$P_n=0.0518(5)$ to find neutral pairs on two neighboring sites.
Here $P_n$ characterize the concentration of interfaces separating
neutral domains (e.g., $\Phi_0^{(h)}$ and $\Phi_2^{(h)}$). Along
these interfaces the evolution is blocked until one of the
invading predators reaches the interface. For $X_s>0$, however,
along these neutral interfaces the site exchange increases the
value of $P_n$ and supports the formation of the well-mixed phase
of neutral species.

\begin{figure}[h]
\begin{center}
             \epsfxsize=7cm
             \epsfbox{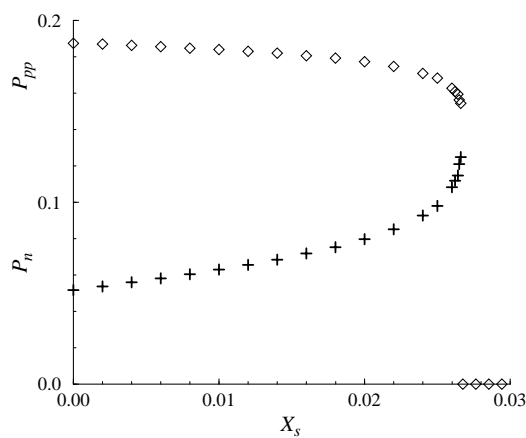}
\end{center}
\caption{Probability of neutral (pluses) and
predator-prey (open diamonds) pairs as a function of $X_s$ for the
four-species cyclic food web (see $a$ in
Fig.~\ref{fig:foodwsub}).}
\label{fig:lv4s1}
\end{figure}

The MC simulations indicate clearly how the value of $P_n$ increases
monotonously with $X_s$ until a threshold value [$X_s < X_{cr}^{(4a)}
=0.02662(2)$]. At the same time the probability $P_{pp}$ of
predator-prey pairs is decreasing as shown in Fig.~\ref{fig:lv4s1}.

Above this threshold value ($X_s > X_{cr}^{(4a)}$) the system
segregates into two types ($\Phi_{02}^{(n)}$ and
$\Phi_{13}^{(n)}$) of growing domains which are mixtures of
neutral pairs. During the domain growing processes these phases
are in contact throughout a wide boundary layer (of type
$\Phi_{0123}^{(c)}$) which serves as a species reservoir and sets
the compositions to be symmetric \cite{szabo:pre04a}. Finally, one
of the two equivalent mixed states will prevail the whole finite
system and then $P_{pp}=0$ and $P_n=0.5$.

Subsystem $b$ represents the situation when the three-species cyclic
defensive alliance ($\Phi_{123}^{(c)}$) is attacked by an external
invader (of type 0). The MC simulations indicate clearly that after some
relaxation process the species 0 dies out and the system remains in the
state $\Phi_{123}^{(c)}$ if $X_s<X_{cr}^{(4b)}=0.0527(1)$. Conversely,
the system develops into the symmetric mixed phase $\Phi_{02}^{(n)}$
($\rho_0=\rho_2=0.5$) for $X_s > X_{cr}^{(4b)}$. In the vicinity of
the transition point the visualization of the species distribution
illustrates the formation of domains of $\Phi_{02}^{(n)}$ whose area
increases (decreases) slowly above (below) the threshold value. The
average velocity of the interfaces (separating the domains of
$\Phi_{02}^{(n)}$ and $\Phi_{123}^{(c)}$)
seems to be proportional to $X_s-X_{cr}^{(4b)}$
as it is observed for another model \cite{szabo:pre04a}. This feature can
explain why the ordering process becomes so slow in the close vicinity
of the transition point.

The four-species subsystem $c$ possesses two equivalent, three-species
cycles that have two common species (1 and 3). In this case
the spatial distribution of species segregates into
two types ($\Phi_{123}^{(c)}$ and $\Phi_{103}^{(c)}$) of growing
domains and finally the whole system is dominated by one of them with
equal probabilities. Notice that the bulk of these phases are not
influenced by the site exchange mechanism whose effect is limited to
the boundaries separating the phases $\Phi_{123}^{(c)}$ and
$\Phi_{103}^{(c)}$. Furthermore the species within the mixed
two-species phase $\Phi_{02}^{(n)}$ are not able to protect each other
because species 1 can invade their territories without restraint.
This is the reason why the domain growing process is not prevented
for $X_s>0$ and the site exchange causes only a slight variation in
the velocity of growing process.

\section{Model A}
\label{sec:ma}

As mentioned above the six-species model $A$ has two equivalent
three-species cycles consisting of disjoint sets of species,
namely $\Phi_{024}^{(c)}$ and $\Phi_{135}^{(c)}$. According to the
simulations one of these phases (with equal probability) will overwhelm
the whole finite system after a domain growing process if the site
exchange probability does not exceeds a threshold value, i.e.,
$X_s < X_{cr}^{(6A)}=0.05592(1)$. On a sufficiently large scale this
domain growing process is similar to those that one can observe for the
kinetic Ising model below the critical temperature if the system is
started from a random initial state (for a survey of domain growth
see \cite{bray:ap94}).

Above the threshold value ($X_s > X_{cr}^{(6A)}$) the system
develops into one of the three (equivalent) two-species states
composed from neutral species. The topological features of the
food web $A$ implies that the phases $\Phi_{03}^{(n)}$,
$\Phi_{14}^{(n)}$, and $\Phi_{25}^{(n)}$ can also be considered as
defensive alliances because their members protect each other
against the external predators in the well-mixed state. During the
transient time one can observe a domain growth with these three
types of domains. Apparently this growth is similar to those
observed in the three-state Potts model below the critical
temperatures \cite{wu:rmp82,grest:prb88}. However, along the
interfaces separating two domains (say $\Phi_{03}^{(n)}$ and
$\Phi_{14}^{(n)}$) there occur a four-species cycle
$\Phi_{0134}^{(c)}$ whose transversal extension is limited by
invasions from both sides. Notice that the corresponding four
species subsystem (see the food web $a$ in
Fig,~\ref{fig:foodwsub}) exhibit a transition from the
four-species cycle ($\Phi_{0134}^{(c)}$ in the mentioned example)
to one of the mentioned neutral pair states at
$X_s=X_{cr}^{(4a)}$. As $X_{cr}^{(6A)} > X_{cr}^{(4a)}$ therefore
the coexistence of the four species is limited to the narrow
strips separating the well-mixed phases of the neutral pairs.

The three-color maps have special points (called three-edge vertices)
where three domains meet. Instead of it, here there are patches
with phases of either $\Phi_{024}^{(c)}$ or $\Phi_{135}^{(c)}$. This
inevitable coexistence of many different phases (during the decomposition
processes) may be the reason why $X_{cr}^{(6A)} > X_{cr}^{(4a)}$.

\section{Model B}
\label{sec:mb}

\begin{figure}[h]
\begin{center}
             \epsfxsize=7cm
             \epsfbox{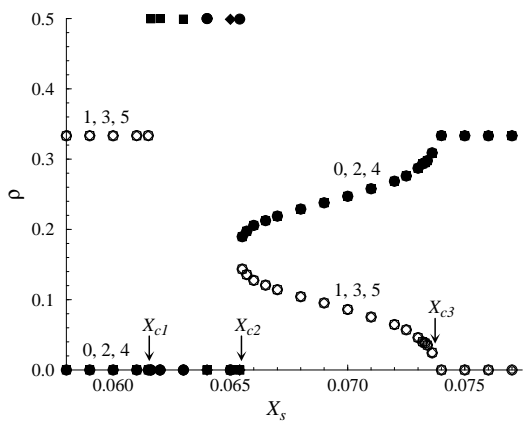}
\end{center}
\caption{Concentration of species {\it vs}. $X_s$. The MC data are denoted by closed circles, open diamonds, closed squares, open circles, closed diamonds, and open squares for the species labelled from 0 to 5. The arrows indicate the
three transition points.}
\label{fig:rho6b}
\end{figure}

In this model a very interesting behavior is found by the MC simulations.
One can observe four different phases and three subsequent phase
transitions when $X_s$ is increased. It is more surprising that the
transition points are very close to each other. To demonstrate the
main features Fig.~\ref{fig:rho6b} illustrates the
average values of species concentrations in the stationary states
reached by the system starting from a random initial state for different
values of $X_s$. Here the open (closed) symbols are used for the species
with odd (even) labels to reduce the confusion coming from the
coincidence of several data. Furthermore, the neutral pairs are denoted
by the same type of symbols.

The cyclic defensive alliance $\Phi_{135}^{(c)}$ is proved to be the
most stable solution if $X_s < X_{c1}^{(6B)}=0.06155(5)$. In the next
region [$X_{c1}^{(6B)} < X_s < X_{c2}^{(6B)}=0.06545(5)$] the final
stationary state will be equivalent to one of the three well-mixed
states of neutral pairs ($\Phi_{03}^{(n)}$, $\Phi_{14}^{(n)}$, and
$\Phi_{25}^{(n)}$). All the six species survive in the third region of
$X_s$ in such a way that $\rho_0=\rho_2=\rho_4$ and $\rho_1=\rho_3=
\rho_5$. If $X_s > X_{c3}^{(6B)}=0.07372(2)$ then only the species
with even labels are present and their coexistence with the same
concentration is maintained by the cyclic invasions (state
$\Phi_{024}^{(c)}$).

\begin{figure}[h]
\begin{center}
             \epsfxsize=7cm
             \epsfbox{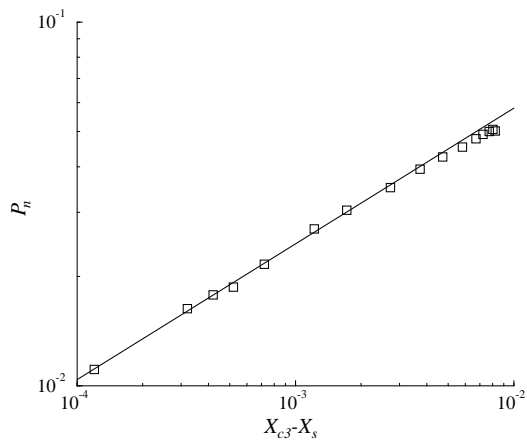}
\end{center}
\caption{Log-log plot of the MC data (open
squares) for the probability of neutral pairs {\it vs}.
$X_{c3}^{(6B)}-X_s$. The fitted power law is denoted by the
straight line with a slope of 0.373.}
\label{fig:betafit}
\end{figure}

The transitions at $X_{c1}$ and $X_{c2}$ can be considered as a first
order transition meanwhile the third one exhibits the features of the
continuous (critical) transitions.
The concentration of the species 1, 3, and 5 vanishes continuously
when approaching $X_{c3}$ from below. More precisely, our data are
consistent with a power law decrease in the close vicinity of the
transition point, that is
$\rho_1=\rho_3=\rho_5 \simeq (X_{c3}-X_s)^{\beta}$ with $\beta=0.37(4)$.
Similar power law behavior (with the same exponent) is found for
$P_n$ as shown in Fig.~\ref{fig:betafit}. At the same time the MC data
show diverging fluctuations
for these quantities that is another relevant feature of the critical
transitions. Unfortunately, we could not deduce an adequate ($\gamma$)
exponent to describe this divergency because of the large statistical
uncertainties in these quantities.

In the presence of absorbing state(s) the extinction of a single
species in the spatial models has very robust features and the
corresponding transition belongs to the directed percolation (DP)
universality class characterized by a higher exponent
($\beta_{DP}=0.58(2)$ for the two-dimensional systems)
\cite{janssen:zpb81,grassberger:zpb82,marro:99,hinrichsen:ap00}.
In the present model, however, three species die out
simultaneously and the background can mediate some interactions
among the vanishing species. Evidently, further analysis is
necessary to quantify the features of this transition as well as
to clarify the reasons causing the distinct behavior.

In contrary to the model $A$ here the formation of one the three
equivalent well-mixed phases of neutral species is limited to a
narrow region [$X_{c1}^{(6B)} < X_s < X_{c2}^{(6B)}=0.06545(5)$].
The formation of these structures differs significantly from those
described for model $A$ because the food web $B$ does not provide
mutual protection for a neutral pair against another neutral pair.
Instead, the well-mixed phases of the neutral pairs destroy
cyclically each other ($\Phi_{03}^{(n)}$ destroys
$\Phi_{14}^{(n)}$ destroys $\Phi_{25}^{(n)}$ destroys
$\Phi_{03}^{(n)}$) via a complicated process. For example, the
offspring of a single species $0$ can occupy the whole territory
of the phase $\Phi_{14}^{(c)}$. As a result, if the growing
domains of the phases $\Phi_{03}^{(c)}$ and $\Phi_{14}^{(c)}$ meet
along a common boundary then the domain  $\Phi_{14}^{(c)}$ is
transformed (via the invasion) into a domain of $\Phi_0^{(h)}$
that is also unstable against the invasions of species $2$ and
$5$.

Within the second region of $X_s$ the six species can coexist for
a sufficiently long time in such a structure where the species
with even labels are in minority. The formation of the mixed
phases of the neutral species begins in those patches where two
species of minorities are missing due to the fluctuations.
According to the previous analysis of the four-species subsystem
$b$ the spontaneous formation of the well-mixed phase of the
corresponding neutral species is permitted because the condition
$X_s > X_{cr}^{(4b)}$ is fulfilled within the given region of
$X_s$. By this means sufficiently large domains of the states
$\Phi_{03}^{(n)}$, $\Phi_{14}^{(n)}$, and $\Phi_{25}^{(n)}$ happen
to form by chance. During their expansion these domains meet one
of their rivals and then one of them is destroyed as described
above.

\begin{figure}[h]
\begin{center}
             \epsfxsize=7cm
             \epsfbox{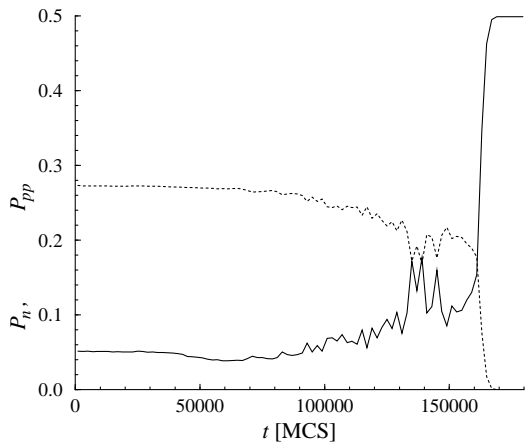}
\end{center}
\caption{Time dependence of the probability of
neutral (dashed line) and predator-prey (solid line) pairs for
$X_s=0.0654$ and $L=2000$. The MC data are smoothed out by
averaging over 2000 MCS.}
\label{fig:x0654t}
\end{figure}

Figure~\ref{fig:x0654t} shows the variation of $P_n(t)$ and
$P_{pp}(t)$ for a given $X_s$ close to the second transition
point. At the beginning these curves are smooth. Later, however,
one can observe larger and larger variations reflecting the
destruction of domains with types $\Phi_{03}^{(n)}$,
$\Phi_{14}^{(n)}$, and $\Phi_{25}^{(n)}$ whose average size
increases monotonously. The series of these larger and larger
invasion processes is ended when only one of these phases prevail
the whole system.

It is conjectured that this (rock-scissors-paper like) cyclic dominance
between the two-species associations plays a crucial role in the
maintenance of the six-species phase.

\section{Model C}
\label{sec:mc}

This model has eight equivalent three-species cyclic associations as
mentioned above. The simulations for a system size of $L=400$
have justified that any of them can be the final stationary state
independently of the value of $X_s$.

\begin{figure}[h]
\begin{center}
             \epsfxsize=8cm
             \epsfbox{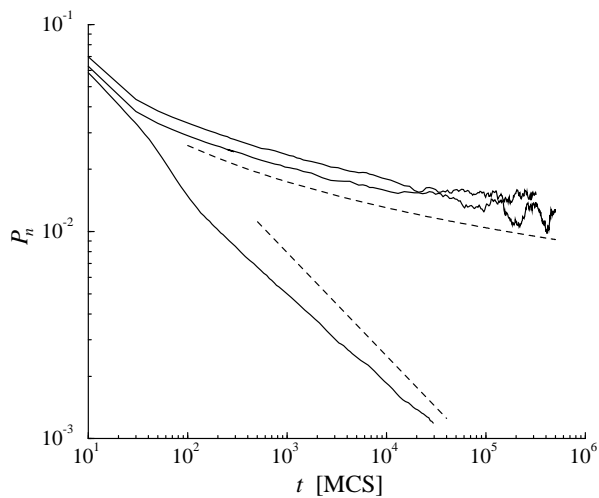}
\end{center}
\caption{Log-log plot of the probability of neutral pairs {\it vs.} time for $X_s=0.05$ (upper solid line) and $X_s=0$ (central solid line). The lower solid line shows the same quantity in the model $A$ for $X_s=0$. To compare the MC results with the theoretical expectations, the upper dashed line represents the
function $P_n(t)=0.12/\ln(t)$ while the lower one shows an algebraic decay, $P_n(t)=0.25/\sqrt{t}$.}
\label{fig:pnt6c}
\end{figure}

It is expected that this "mono-domain" state is developed via a
domain growing process. This phenomenon can be described by the
variation of the probability of neutral pairs vanishing in the
bulk of "mono-domain" states. The results of MC simulations (for
$L=2500$) are plotted in Fig.~\ref{fig:pnt6c} where the data are
averaged over a sampling time increased from 2 to 2000 MCS.

Figure~\ref{fig:pnt6c} shows that the very slow domain growing process
for $X_s=0$ becomes even slower for $X_s=0.05$. Indeed we cannot exclude
the blocking of coarsening process as it happens in the voter
model for higher dimensions ($d>2$) \cite{ben-naim:pre96}.
Keeping in mind the large statistical uncertainties (for large times) these data remind us
to the growing processes observed for the two-dimensional voter model
where the concentration of domain walls vanishes as $1/\ln(t)$ plotted
also in Fig.~\ref{fig:pnt6c} \cite{liggett:85,ben-naim:pre96}.
The voter model exemplifies the coarsening process in a broad class
of models undergoing a phase ordering without surface tension
\cite{dornic:prl01}. In the present model the definition of the
boundary separating two cyclic three-species phases is confusing because
these phases may involve one or two common species (compare
$\Phi_{015}^{(c)}$ with $\Phi_{012}^{(c)}$ or $\Phi_{123}^{(c)}$).

Strikingly different growth process is found for model $A$
(discussed previously) where the system has only two cyclic
three-species phases composed from
disjoint sets of species ($\Phi_{024}^{(c)}$ or $\Phi_{135}^{(c)}$).
In order to visualize the relevant differences Fig.~\ref{fig:pnt6c}
shows also the MC results obtained in the model $A$ for $X_s=0$ and
$L=4000$. These numerical results indicate that here the asymptotic
behavior of $P_n(t)$ is close to the theoretical prediction
($P_n(t) \sim 1/\sqrt{t}$) characterizing the growth controlled by
the reduction of an interfacial energy \cite{bray:ap94,grest:prb88}.

Evidently, further research is required to clarify those
ingredients of these models that are responsible for the different
growth processes. Furthermore, here it is worth mentioning that
the logarithmic growth supports the maintenance of biodiversity
for a very long time if the system is sufficiently large.

\section{Model D}
\label{sec:md}

\begin{figure}[h]
\begin{center}
             \epsfxsize=7.5cm
             \epsfbox{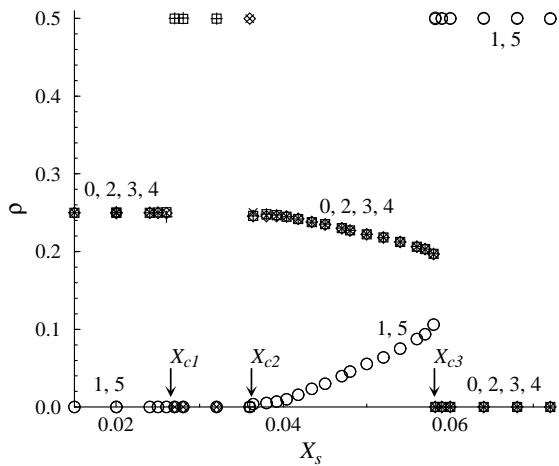}
\end{center}
\caption{Concentration of species {\it vs}. $X_s$ in the model $D$. The MC data are denoted by symbols with fourfold symmetry (pluses, Xs, open squares, and diamonds) for those species belonging to four-species cyclic defensive alliance, and by open circles for the species $1$ and $5$. The arrows indicates
the critical values of $X_s$ where transitions appear.}
\label{fig:rho6d}
\end{figure}

For this food web the system has a four-species cyclic defensive
alliance ($\Phi_{0234}^{(c)}$ as shown in Fig.~\ref{fig:foodweb})
that is stable if $X_s=0$ \cite{szabo:pre01}. According to the
simulations for weak rate of site exchange the species $1$ and $5$
die out within a short transient time. In the corresponding
four-species subsystem (see Sect.~\ref{sec:4s}) the cyclic
invasions can sustain their coexistence if $X_s < X_{cr}^{(4a)}$.
For higher $X_s$ the spatial distribution of these four species
decomposes into two types of two-species domains
($\Phi_{03}^{(n)}$ and $\Phi_{24}^{(n)}$) as described above. If
$X_s$ is increased then this process becomes faster, and above a
threshold value the domains of phases $\Phi_{03}^{(n)}$ and
$\Phi_{24}^{(n)}$ are built up by these processes before their
mortal enemy (species $1$ and $5$) would die out. One can easily
check that the species $1$ ($5$) can occupy the whole territory of
the phases $\Phi_{03}^{(n)}$ ($\Phi_{24}^{(n)}$). Thus, for a
sufficiently intensive mixing the species $1$ and $5$ will have an
enhanced chance to survive because they are fed by the emerging
domains of $\Phi_{03}^{(n)}$ and $\Phi_{24}^{(n)}$. For large
$X_s$ this mechanism can be so effective that finally the phase
$\Phi_{15}^{(n)}$ will dominate the whole system as shown in
Fig.~\ref{fig:rho6d}. Notice that the well mixed phase
$\Phi_{15}^{(n)}$ is also a defensive alliance because species $1$
and $5$ protect each other against the different invaders.

Figure~\ref{fig:rho6d} shows that four types of final stationary
states can be distinguished when $X_s$ is increased. In the first
region ($X_s < X_{c1}^{(6D)}=X_{cr}^{(4a)}$) the species of the
phase $\Phi_{0234}^{(c)}$ take place with the same concentration
meanwhile $P_n$ and $P_{pp}$ vary with $X_s$ as plotted in
Fig.~\ref{fig:lv4s1}. Starting from a random initial state this
system evolves into one of the neutral-pair phases
$\Phi_{03}^{(n)}$ and $\Phi_{24}^{(n)}$ if $X_{c1}^{(6D)} < X_s <
X_{c2}^{(6D)}$ where the value of $X_{c2}^{(6D)}$ depends on the
size $L$ as it will be discussed below. Within the third region
($X_{c2}^{(6D)} < X_s < X_{c3}^{(6D)}$) the six species coexist in
such a way that $\rho_0=\rho_2=\rho_3=\rho_4 > \rho_1=\rho_5 > 0$.
If $X_s > X_{c3}^{(6D)}=0.0581(1)$ then only the species $1$ and
$5$ will survive with forming a phase $\Phi_{15}^{(n)}$.

\begin{figure}[h]
\begin{center}
             \epsfxsize=7cm
             \epsfbox{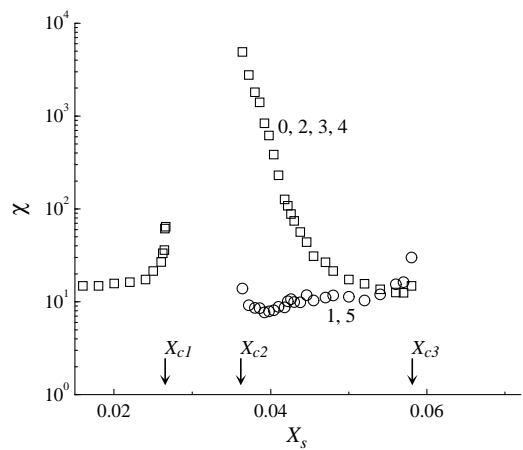}
\end{center}
\caption{Fluctuation of concentration as a function of $X_s$ for the species as denoted by the figures.}
\label{fig:chi6d}
\end{figure}

In this model the first and the third transitions exhibit a sudden
change both in concentrations and in pair probabilities
($P_n$ and $P_{pp}$). However, the classification of the second
transition (as well as the numerical value of $X_{c2}^{(6D)}$) is
difficult because of its unusual features. Namely, $\rho_1$ and
$\rho_5$ decrease monotonously when $X_s$ is decreased within the
third region. At the same time we have observed a dramatic
increase in the fluctuation of concentration for the majority
species (see Fig.~\ref{fig:chi6d}) meanwhile the corresponding
average concentrations are close to 1/4 (see
Fig.~\ref{fig:rho6d}). To avoid the undesired effects of these
fluctuations the simulations were performed for such a large
system sizes (the linear size is increased up to $L=2600$) where
the current value of concentrations were significantly larger than
the amplitude of fluctuation characterized by their standard
deviation dependent on size ($\sqrt{\chi / N}$). For smaller sizes
the system evolves into the phase $\Phi_{03}^{(n)}$ or
$\Phi_{24}^{(n)}$.

Figure~\ref{fig:chi6d} demonstrates clearly the absence of
fluctuations for the missing species as well as for the phases
composed of neutral species. The MC data refer to an exponential
increase in $\chi_0$, $\chi_2$, $\chi_3$, and $\chi_4$ if $X_s$
approaches $X_{c2}$ from above.

Within the region of large fluctuations the time dependence of
concentrations shows the occurrence and destruction of domains
with the structure $\Phi_{03}^{(n)}$ and $\Phi_{24}^{(n)}$. These
domains are invaded by the sparsely dispersed species $1$ and $5$.
The resultant phase, however, is not stable at the given $X_s$ and
sooner or later the phases $\Phi_{03}^{(n)}$ or $\Phi_{24}^{(n)}$
are formed again. In some sense the evolution of the
spatio-temporal pattern is similar to those described by the
mentioned spatial rock-scissors-paper games
\cite{tainaka:prl89,frachebourg:jpa98,szabo:pre02a} as well as in
the forest-fire models
\cite{bak:pla90,drossel:prl92,grassberger:jpa93}. In the present
case two cycles interfere in a complex way because cyclic
dominance emerges between the different set of associations and
this mechanism supports the maintenance of biodiversity with more
associations (and species).

Due to the mentioned difficulties we couldn't determine accurately how
concentrations $\rho_1$ and $\rho_5$ vanish when approaching
$X_{c2}^{(6D)}$ from above. In the light of the MC results one can
conclude a weakly first order phase transition (sudden jump in
$\rho_1=\rho_5$) accompanied with an increase in $\chi_1=\chi_5$
as illustrated in Fig.~\ref{fig:chi6d}.

\section{Conclusions}
\label{sec:conc}

The numerical analysis of the four different six-species, spatial
predator-prey models has highlighted the large variety of states and phenomena (transitions) that can occur in self-sustaining, multi-species ecological systems. Due to their simplicity the present models have allowed us to consider quantitatively some interesting phenomena. Instead of recalling the curious features found for these oversimplified models henceforth I wish to concentrate on those general messages concluded from these examples.  

One of the relevant messages is related to the increase in the degree of freedom. Many recent works are focused on spatial systems with one, two, or three species. For larger number of species the present models exemplify how the simple dynamical rules build up more complex objects
(alliances) that can be considered as additional species
participating in the evolutionary competition. Shortly, the degree
of freedom (here the number of species) is increased by the
spontaneously occurring alliances having particular composition
and spatial structure. One can believe that the food web of these
systems should be extended by these complex objects (with a
suitable invasion rates) to have a better picture about the whole
spatial system.

In the context of game theory (for details see the refs.
\cite{maynard:82,weibull:95,hofbauer:98,gintis:00,miekisz:jpa04}),
the interaction of species is described by the payoff matrix (the
adjacency matrix of the food web in the present models) that
determines the Nash equilibrium from which the unilateral
deviation yields loss. For many evolutionary (dynamical) rules the
Nash equilibrium represents an evolutionary stable state under the
mean-field conditions. The Nash equilibrium is not necessary
unique and it can be a mixed state (strategy). In the spatial
models with short range interactions several species can be absent
from a given territory and the Nash equilibrium for the
corresponding subsystem can be distinct from those characterizing
the complete system. The noises affect the stability of these
spatial states \cite{miekisz:jsp04} and modify the average
velocity of invasions along the interfaces separating the
competing phases \cite{szabo:pre04a}. The above results justify
that the invasion processes between the ``stable'' states of
subsystems can play a crucial role in the formation of the final
stationary states.

For many cases the final stationary state is reached through a
coarsening process where the alliances (degenerated Nash equilibria)
form growing domains. The corresponding domain growth process is
affected by the additional phases (and mechanisms) along interfaces.
It is not yet known how these effects modify the growth process
yielding significantly slower (logarithmical) domain growth.

Most of the (first order) transitions occur at such a value of
$X_s$ where the invasion rate between two dominant alliances
vanishes as detailed previously \cite{szabo:pre04a}. Remarkable
exceptions are the fourth transition in model $B$ and the second
one in model $D$. In both cases more than one species die out
simultaneously and their extinction is accompanied by an enhanced
concentration fluctuation for the surviving species and the
resultant spatio-temporal structures can mediate unusual
interactions between the vanishing species. These transitions are
distinct inherently and their classification requires further
time-consuming numerical investigations.

The spontaneous formation of complex spatio-temporal structures is
one of the most interesting predictions of these simple models
because it sustains an abundance of species. Furthermore, this
feature mixes the concept of the part and whole. At the same time
the corresponding system can be interpreted as a set of objects
that create objects in a way characterizing the living systems
\cite{crutchfield:nl04}. In the light of the present results one
can easily imagine that the choice of more complex food webs can
result in the emergence of more and more complex objects (e.g.,
alliances of alliances with particular spatio-temporal
structures). Besides it these models imply the possibility that
the evolution of food web by mutations
\cite{drossel:jtb01,holt:er02,chowdhury:pre03} and the evolution
of the spatial ecological system are strongly related to each
other while the spatial inhomogeneity can be enhanced in the
system.

\section{Acknowledgment}
Supports from the Hungarian National Research Fund (T-47003) and the European Science Foundation (COST P10) are acknowledged.


\Bibliography{99}
\bibitem{maynard:82}Maynard~Smith J 1982 {\it Evolution and the theory of games} (Cambridge University Press, Cambridge)

\bibitem{weibull:95}Weibull J W 1995 {\it Evolutionary Game Theory}(MIT Press, Cambridge, Mass.)

\bibitem{hofbauer:98}Hofbauer J and Sigmund K 1998 {\it Evolutionary Games
and Population Dynamics} (Cambridge University Press, Cambridge)

\bibitem{gintis:00}Gintis H 2000 {\it Game Theory Evolving} (Princeton
  University Press, Princeton)

\bibitem{drossel:ap01}Drossel B 2001 {\it Adv. Phys.} {\bf 50} 209

\bibitem{miekisz:jpa04}Miekisz J 2004 \JPA {\bf 37} 9891

\bibitem{pekalski_cse04}Pekalski A 2004 {\it Comp. Sci. Eng.} {\bf 6} 62

\bibitem{he_ijmp05}He M F, Pan Q H and Wang S 2005 {\it Int. J. Mod. Phys. C} {\bf 16} 177

\bibitem{watt:je47}Watt A S 1947 {\it J. Ecol.} {\bf 35} 1

\bibitem{rasmussen:science04}Rasmussen S \etal 2004 {\it Science} {\bf 303} 963

\bibitem{tainaka:prl89}Tainaka K 1989 \PRL {\bf 63} 2688

\bibitem{sato:mmit97}Sato K, Konno N and Yamaguchi T 1997 {\it Mem. Muroran Inst. Tech.} {\bf 47} 109

\bibitem{frachebourg:jpa98}Frachebourg L and Krapivsky P L 1998 \JPA {\bf 31} L267

\bibitem{tainaka:pla93}Tainaka K 1993 \PL A {\bf 176} 303

\bibitem{frean:prs01}Frean M and Abraham E D 2001 {\it Proc. Roy. Soc. Lond.} B {\bf 268} 1

\bibitem{tainaka:epl91}Tainaka K and Itoh Y 1991 {\it Europhys. Lett.} {\bf 15} 399

\bibitem{szabo:pre02a}Szab{\'o} G and Szolnoki A 2002 \PR E {\bf 65} 036115

\bibitem{boerlijst:pd91}Boerlijst M C and Hogeweg P 1991 {\it Physica} D {\bf 48} 17

\bibitem{szabo:pre01}Szab{\'o} G and Cz{\'a}r{\'a}n T 2001 \PR E {\bf 64} 042902

\bibitem{sato:amc02}Sato K, Yoshida N and Konno N 2002 {\it Appl. Math. Comp.} {\bf 126} 255

\bibitem{szabo:pre04a}Szab{\'o} G and Sznaider G A 2004 \PR E {\bf 69} 031911

\bibitem{ravasz:pre04}Ravasz M, Szab\'o G and Szolnoki A 2004 \PR E {\bf 70} 012901

\bibitem{traulsen:pre03}Traulsen A and Schuster H G 2003 \PR E {\bf 68} 046129

\bibitem{traulsen:pre04}Traulsen A and Claussen J C 2004 \PR E {\bf 70} 046128

\bibitem{bray:ap94}Bray A J 1994 {\it Adv. Phys.} {\bf 43} 357

\bibitem{wu:rmp82}Wu F Y 1982 \RMP {\bf 54} 235

\bibitem{grest:prb88}Grest G S, Anderson M P and Srolovitz D J 1988 \PR B {\bf 38} 4752

\bibitem{janssen:zpb81} Janssen H K 1981 {\it Z. Phys. B Cond. Mat.} {\bf 42} 151

\bibitem{grassberger:zpb82}Grassberger P 1982 {\it Z. Phys. B Cond. Mat.} {\bf 47} 365

\bibitem{marro:99}Marro J and Dickman R 1999 {\it Nonequilibrium Phase Transitions in Lattice Models} (Cambridge University Press, Cambridge)

\bibitem{hinrichsen:ap00}Hinrichsen H 2000 {\it Adv. Phys.} {\bf 49} 815

\bibitem{ben-naim:pre96}Ben-Naim E, Frachebourg L and Krapivsky P L 1996 \PR E {\bf 53} 3078

\bibitem{liggett:85}Liggett T M 1985 {\it Interacting Particle Systems} (Springer-Verlag, New York)

\bibitem{dornic:prl01}Dornic I \etal 2001 \PRL {\bf 87} 045701

\bibitem{bak:pla90}Bak P, Chen K and Tang C 1990 \PL A {\bf 147} 297

\bibitem{drossel:prl92}Drossel B and Schwabl F 1993 \PRL {\bf 69} 1629

\bibitem{grassberger:jpa93}Grassberger P 1993 \JPA {\bf 26} 1081

\bibitem{miekisz:jsp04}Miekisz J 2004 {\it J. Stat. Phys.} {\bf 117} 99

\bibitem{crutchfield:nl04}Crutchfield J P and G{\"o}rnerup O 2004 nlin.AO/0406058

\bibitem{drossel:jtb01}Drossel B, Higgs P G and McKane A J 2001 {\it J. Theor. Biol.} {\bf 208} 91

\bibitem{holt:er02}Holt R D 2002 {\it Ecol. Res.} {\bf 17} 261

\bibitem{chowdhury:pre03}Chowdhury D and Stauffer D 2003 \PR E {\bf 68} 041901


\endbib
\end{document}